\documentclass[twocolumn,groupedaddress]{revtex4-2}
\usepackage{float}
\usepackage{tikz}
\usetikzlibrary{quantikz2}
\definecolor{wred}{HTML}{FC2855}
\definecolor{worange}{HTML}{FD8F2A}
\definecolor{wgreen}{HTML}{15A94E}
\definecolor{wblue}{HTML}{3D8EDD}
\definecolor{wpurple}{HTML}{DF6DF7}

\begin{document}

%\preprint{}

\title{Halving the Cost of Controlled Time Evolution}

\author{William A. Simon}
\email{Contact author: william.andrew.simon@gmail.com}
\affiliation{Department of Physics and Astronomy, Tufts University.}
 
\author{Peter J. Love}%
\affiliation{Department of Physics and Astronomy, Tufts University.}

\date{\today}

\begin{abstract}
Quantum simulation is a promising application for quantum computing.
Quantum simulation algorithms may require the ability to control the time evolution unitary.
Naive techniques to control a unitary can substantially increase the required computational resources.
A standard approach to controlling Trotterized time evolution doubles the number of single-qubit arbitrary rotations.
Here, we describe a compilation scheme that does not increase the number of arbitrary rotations for symmetric Trotterizations, which applies to second-order and higher Suzuki-Trotter decompositions.
This halves the number of arbitrary rotations required to implement controlled, Trotterized time evolution compared to the standard approach. 
Arbitrary rotations contribute significantly to resource estimates in a fault-tolerant architecture due to the number of required magic states.
Therefore, arbitrary rotations dominate the $T$-cost of fault-tolerant implementations of quantum simulation. 
This construction reduces the number of arbitrary rotations for controlled Trotter evolution to that of uncontrolled Trotter evolution, thereby reducing the cost of fault-tolerant quantum simulation.
\end{abstract}

\maketitle

\section{\label{sec:intro}Introduction}

The simulation of quantum systems was the original motivating idea behind quantum computers \cite{feynman1982simulating, manin2007mathematics}.
Despite the discovery of other applications of quantum computers \cite{shor1999polynomial,abrams1999quantum,harrow2009quantum}, quantum simulation remains an exciting prospect for exponential quantum advantage on both near \cite{google2020hartree,evered2025probing,otoc} and far-term quantum devices \cite{aspuru2005simulated,jordan2012quantum}.

The number of operations required by an algorithm determines the error rate that needs to be achieved.
In a fault-tolerant quantum architecture, quantum error correction is used to reduce the logical error rate \cite{gottesman1997stabilizer,fowler2012surface,litinski2019game}.
As the target error rate becomes lower, more physical qubits are needed to encode a single logical qubit. 
Therefore, reduing the number of operations required by an algorithm not only reduces the time it takes, but it allows these algorithms to be run on earlier, smaller devices.

Numerical resource estimates help determine how large a quantum computer must be to solve certain problems.
Under the surface code \cite{fowler2012surface, litinski2019game}, the non-transversal $T$ gate is the most expensive operation to perform due to the cost of preparing and consuming a magic state \cite{bravyi2005universal, gidney2024magic}.
In fault-tolerant implementations, arbitrary rotations must be decomposed into a discrete gate set using the Solovay-Kitaev algorithm \cite{Kitaev_1997, kitaev2002classical, dawson2005solovay} or other techniques \cite{gidney2018halving, ross_selinger, bocharov_rus, Kliuchnikov_2013, Kliuchnikov2023shorterquantum, kim2025catalytic}.
These techniques use multiple $T$ gates to approximate each arbitrary rotation.
This means arbitrary rotations of logical qubits often require the most resources in fault-tolerant quantum simulation algorithms.

A fundamental operation in quantum simulation is the time evolution unitary generated by the Hamiltonian of a quantum system \cite{lloyd1996universal}.
Trotterization \cite{lie1893theorie, trotter1958approximation, trotter1959product, suzuki1976generalized,suzuki1985decomposition, suzuki1986quantum, suzuki1990fractal, hatano2005finding, suzuki2012quantum} refers to several techniques to efficiently approximate the time evolution operator.
Quantum simulation algorithms may also require \textit{controlled} applications of the time evolution unitary.
Controlling a Trotter decomposition through standard techniques doubles the number of arbitrary rotations.

In this work, we describe a compilation scheme for controlling Trotter decompositions that does not increase the number of arbitrary rotations.
Our techniques can be applied to symmetric Trotter decompositions, which are found in second-order and higher Suzuki-Trotter approximations \cite{suzuki1976generalized,suzuki1985decomposition, suzuki1986quantum, suzuki1990fractal, hatano2005finding, suzuki2012quantum}.
Compared to the standard technique, this construction halves the number of arbitrary rotations required to implement controlled time evolution.
Arbitrary rotations lead to large $T$-counts in fault-tolerant architectures since each arbitrary rotation of a logical qubit consumes multiple magic states.
By reducing the total number of arbitrary rotations by a factor of two, we therefore improve the dominant cost of controlled time evolution.

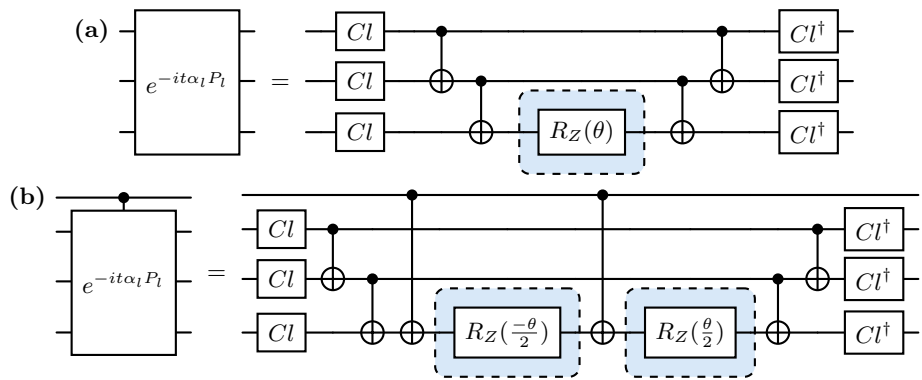
\begin{figure*}[t]
    \centering
    \begin{quantikz}[column sep=0.2cm, row sep=0.1cm, font=\small]
        \lstick{\textbf{(a)}} & \gate[3]{e^{-it\alpha_lP_l}} & \\
        & \ghost{Cl^\dagger}           & \\
        & \ghost{R_Z(\theta)}          &
    \end{quantikz}
    =
    \begin{quantikz}[column sep=0.2cm, row sep=0.1cm, font=\small]
        && \gate{Cl} && &  \ctrl{1} &     &    & &  &  &  &    &   \ctrl{1} && & \gate{Cl^\dagger}  & \\
        && \gate{Cl} && & \targ{} & \ctrl{1} && &                  & & & \ctrl{1} & \targ{} & && \gate{Cl^\dagger} & \\
        && \gate{Cl} && &        & \targ{} & && \gate{R_Z(\theta)} \gategroup[1,steps=1, style={dashed, rounded corners, fill=wblue!20}, background]{} && & \targ{} &         & && \gate{Cl^\dagger} &
    \end{quantikz}

    \begin{quantikz}[column sep=0.2cm, row sep=0.1cm, font=\small]
        \lstick{\textbf{(b)}} & \ctrl{1} &  \\
        & \gate[3]{e^{-it\alpha_lP_l}} & \\
        & \ghost{Cl^\dagger}           & \\
        & \ghost{R_Z(\theta)}          &
    \end{quantikz}
    =
    \begin{quantikz}[column sep=0.2cm, row sep=0.1cm, font=\small]
        & & &         & \ctrl{3} &  &   &              &  \ctrl{3}  &&       &&  &&   & \\
        & \gate{Cl} & \ctrl{1} &        & && &  &   &              &  &       & & \ctrl{1} &\gate{Cl^\dagger}  & \\
        & \gate{Cl} &\targ{} & \ctrl{1} & &&     & &             & & && \ctrl{1} & \targ{} &\gate{Cl^\dagger} & \\
        & \gate{Cl} &       & \targ{} & \targ{} & &\gate{R_Z(\frac{-\theta}{2})} \gategroup[1,steps=1, style={dashed, rounded corners, fill=wblue!20}, background]{}& & \targ{} & &\gate{R_Z(\frac{\theta}{2})} \gategroup[1,steps=1, style={dashed, rounded corners, fill=wblue!20}, background]{} && \targ{} &         &\gate{Cl^\dagger} &
    \end{quantikz}
    \caption{
        \label{fig:pauli-evolution}
        \textbf{Pauli Operator Time Evolution.}
        \textbf{(a)} The general structure of a circuit implementing the time evolution unitary generated by a $k$-local Pauli operator is shown.
        \textbf{(b)} The standard circuit for implementing the \textit{controlled} time evolution unitary generated by a $k$-local Pauli operator is shown. This circuit requires two arbitrary rotations.
    }
\end{figure*}

\section{\label{sec:trotterization}Trotterization}

Trotterization applies to Hamiltonians that can be efficiently expressed as a linear combination of operators: 
\begin{equation}
    H = \sum_{l=0}^{L-1} \alpha_l O_l
\end{equation}
where $\alpha_l$ is the coefficient of the term $O_l$.
Using Trotterization in digital quantum simulation algorithms requires quantum circuits that efficiently implement the time evolution unitary generated by each operator ($e^{-itO_l}$).

The first-order ($p=1$) Suzuki-Trotter decomposition is given by:
\begin{equation}
    \label{eq:first-order}
    U_{1}(t) \equiv \prod_{l=0}^{L-1} e^{-i \alpha_l t O_l}
\end{equation}
The second-order ($p=2$) Suzuki-Trotter decomposition is defined by \cite{suzuki1985decomposition,suzuki1986quantum,suzuki2012quantum}:
\begin{equation}
    \label{eq:second-order}
    \begin{split}
        U_{2}(t) &\equiv \prod_{l=L-1}^{0} e^{-i\frac{\alpha_lt}{2} O_l} \prod_{l=0}^{L-1} e^{-i\frac{\alpha_lt}{2} O_l} \\
        &= U_1^\prime(t / 2) U_1(t / 2)
    \end{split}
\end{equation}
where $U_1^\prime$ is defined by a first-order decomposition, but with the order of the terms in the product reversed.
The two neighboring time evolutions of the operator $O_{L-1}$ can be merged together.
We omit this optimization for clarity, yet it can be applied to the compilation techniques we describe below.

Higher-order approximations can be constructed recursively from the second-order decomposition following Suzuki et al. \cite{suzuki1990fractal}:
\begin{equation}
    \label{eq:higher-order}
    U_{p}(t) = \big[U_{p^*}(\alpha_{p} t)\big]^2 U_{p^*}((1-4\alpha_{p}) t) \big[U_{p^*}(\alpha_{p} t)\big]^2
\end{equation}
where $p^* = p - 2$ and $\alpha_{p}$ is given by:
\begin{equation}
    \alpha_{p} = (4 - 4^{1/(p-1)})^{-1}
\end{equation}

In this work, we focus on Hamiltonians defined by a linear combination of $k$-local Pauli operators ($P_l$).
Circuits implementing the time evolution generated by a $k$-local Pauli operator can be constructed efficiently (subfigure \ref{fig:pauli-evolution}a) \cite{nielsen2001quantum}.
The only non-Clifford operation in these circuits is the arbitrary rotation, making it the most expensive component in a fault-tolerant architecture.
A first-order Trotter decomposition requires $L$ arbitrary rotations.
A second-order Trotter decomposition requires $2L$ arbitrary rotations.
Higher-order Trotter decompositions require $2L 5^{(p/2) - 1}$ arbitrary rotations.

Controlling the time evolution of a Pauli operator can be accomplished by controlling the arbitrary rotation, which requires two arbitrary rotations (subfigure \ref{fig:pauli-evolution}b).
Using this method to control a Trotter decomposition results in twice the number of arbitrary rotations as compared to the uncontrolled circuit.
The desired angle of these rotations is also halved, and smaller angle rotations may require more magic states to achieve the same overall error.

\section{\label{sec:methods}Optimized Compilation}

\begin{figure*}[t]
    \centering
    \begin{quantikz}[column sep=0.2cm, row sep=0.1cm, font=\small]
        \lstick{\textbf{(a)}} & \gate{} & \\
        & \gate[3]{e^{\mp it\alpha_lP_l}}\wire[u]{q} & \\
        & \ghost{Cl^\dagger}           &\\
        & \ghost{R_Z(\theta)}          &
    \end{quantikz}
    $\equiv$
    \begin{quantikz}[column sep=0.2cm, row sep=0.1cm, font=\small]
        \ghost{} & \ctrl[open]{1} & \ctrl{1} & \\
        & \gate[3]{e^{it\alpha_lP_l}} &  \gate[3]{e^{-it\alpha_lP_l}} &\\
        & \ghost{Cl^\dagger}           & &\\
        & \ghost{R_Z(\theta)}          & &
    \end{quantikz}
    =
    \begin{quantikz}[column sep=0.2cm, row sep=0.1cm, font=\small]
        \ghost{} &  &  &         & \ctrl[open]{3} &  &                   &&  \ctrl[open]{3}  &    &  &  & \\
        & \gate{Cl} &  \ctrl{1} &         & &      &&             &  &       & \ctrl{1} & \gate{Cl^\dagger}  & \\
        & \gate{Cl} & \targ{} & \ctrl{1} & &    &&               & & \ctrl{1} & \targ{} & \gate{Cl^\dagger} & \\
        & \gate{Cl} &        & \targ{} & \targ{} & & \gate{R_Z(\theta)} \gategroup[1,steps=1, style={dashed, rounded corners, fill=wblue!20}, background]{} && \targ{} & \targ{} &         & \gate{Cl^\dagger} &
    \end{quantikz}

    \begin{quantikz}[column sep=0.2cm, row sep=0.5cm, font=\small]
        \lstick{\textbf{(b)}} &&& \gate{} & \\
        & \qwbundle{N} && \gate[3]{U_2(\pm t)} \wire[u]{q} &
    \end{quantikz}
    =
    \begin{quantikz}[column sep=0.2cm, row sep=0.5cm, font=\small]
        &&& \gate{} & \hdots{} & \gate{} & \hdots{} & \gate{} & \\
        & \qwbundle{N} && \gate[3]{e^{\mp i\frac{t}{2}\alpha_0P_0}} \wire[u]{q} &\hdots{} & \gate[3]{e^{\mp it\alpha_{L-1}P_{L-1}}} \wire[u]{q} & \hdots{} & \gate[3]{e^{\mp i\frac{t}{2}\alpha_0P_0}} \wire[u]{q} &
    \end{quantikz}
    \caption{
        \label{fig:directional-evolution}
        \textbf{\textit{Directionally-Controlled} Time Evolution.}
        \textbf{(a)} The circuit for directionally controlling the time evolution operator generated by a multi-qubit Pauli operator only requires two additional CNOT gates compared to the uncontrolled operation. 
        \textbf{(b)} For symmetric Trotterizations, directional control can be implemented by directionally controlling the time evolution of each term. 
    }
\end{figure*}
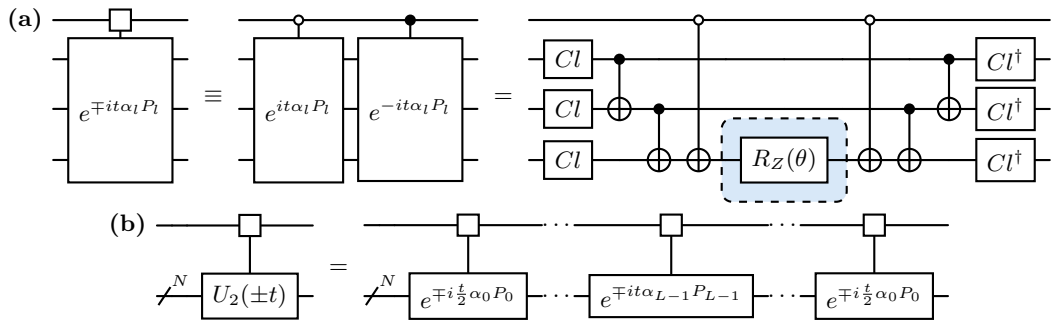

Wecker et al. \cite{wecker2015solving} introduce a techinque referred to as ``directional control" \cite{kivlichan2020improved}.
The aim of directionally controlled time evolution is to evolve the system forward in time if the control qubit is in the $\ket{1}$ state and backwards in time if the control is in the $\ket{0}$ state:
\begin{equation}
    U(\pm t) \equiv \ket{0}\bra{0} \otimes e^{itH} + \ket{1}\bra{1} \otimes e^{-itH}
\end{equation}
Directional evolution of a Pauli operator ($e^{\mp i t \alpha_l P_l}$) can be achieved by conjugating the arbitrary rotation with two $\ket{0}$-controlled CNOT gates (subfigure \ref{fig:directional-evolution}a) \cite{wecker2015solving}.

Symmetric Trotter decompositions can be directionally-controlled ($U_p(\pm t)$) by directionally-controlling the time evolutions of each term (subfigure \ref{fig:directional-evolution}b).
The first-order decomposition (Eq. \ref{eq:first-order}) is not symmetric; directionally controlling the time evolution of each term in the first-order decomposition would result in evolving the system under \textit{different} effective Hamiltonians for the forward and reverse directions \cite{kivlichan2020improved}. 

When the directionally controlled unitary is applied to an eigenstate, it doubles the effect of phase kickback.
Wecker et al. \cite{wecker2015solving} and Kivlichan et al. \cite{kivlichan2020improved} use this fact to reduce resource estimates for eigenstate energy estimation by using the directionally controlled unitary in Quantum Phase Estimation. 

\begin{figure*}[t]
    \centering
    \begin{quantikz}[column sep=0.2cm, row sep=0.5cm, font=\small]
        \lstick{\textbf{(a)}}&&&  \ctrl{1} & \ghost{} \\
        & \qwbundle{N} && \gate[3]{U_2(t)} & \ghost[3]{U_1^\dagger(t / 2)} 
    \end{quantikz}
    =
    \begin{quantikz}[column sep=0.2cm, row sep=0.5cm, font=\small]
        && \ctrl{1} \gategroup[2,steps=2, style={dashed, rounded corners, fill=wblue!20}, background]{{\sc \footnotesize{4L Rotations}}} & \ctrl{1} && \ghost{} \\
        && \gate[3]{U_1(t / 2)} & \gate[3]{U_1^\prime(t / 2)}\wire[u]{q}  &&
    \end{quantikz}
    =
    \begin{quantikz}[column sep=0.2cm, row sep=0.5cm, font=\small]
        && \gategroup[2,steps=2, style={dashed, rounded corners, fill=wblue!20}, background]{{\sc \footnotesize{2L Rotations}}} & \gate{} && \\
        && \gate[3]{U_1(t / 2)} & \gate[3]{U_1^\prime(\pm t / 2)}\wire[u]{q}  &&
    \end{quantikz}

    \begin{quantikz}[column sep=0.2cm, row sep=0.5cm, font=\small]
        \lstick{\textbf{(b)}}&&&  \ctrl{1} & \ghost{} \\
        & \qwbundle{N} && \gate[3]{U_{p}(t)} & \ghost[3]{U_1^\dagger(t / 2)}
    \end{quantikz}
    =
    \begin{quantikz}[column sep=0.2cm, row sep=0.5cm, font=\small]
        & & & \ctrl{1} & \gate{} & \gate{} & \ghost{} \\
        & \gate[3]{U_{p^*}(\alpha_p t)} & \gate[3]{U_{p^*}(\alpha_p t)}  & \gate[3]{U_{p^*}((1-4\alpha_p) t)} & \gate[3]{U_{p^*}(\pm \alpha_p t)} \wire[u]{q} &\gate[3]{U_{p^*}(\pm \alpha_p t)} \wire[u]{q}  &
    \end{quantikz}
    \caption{
        \label{fig:improved-compilation}
        \textbf{\textit{Controlled} Time Evolution.}
        \textbf{(a)} The controlled time evolution of a second-order Suzuki-Trotter decomposition can be achieved using only $2L$ rotations, whereas the naive construction shown in Figure~\ref{fig:pauli-evolution} requires $4L$ rotations.
        \textbf{(b)} Controlled time evolution for higher-order Suzuki-Trotter decompositions can be implemented with fewer CNOT gates by using uncontrolled and directionally controlled components.
    }
\end{figure*}
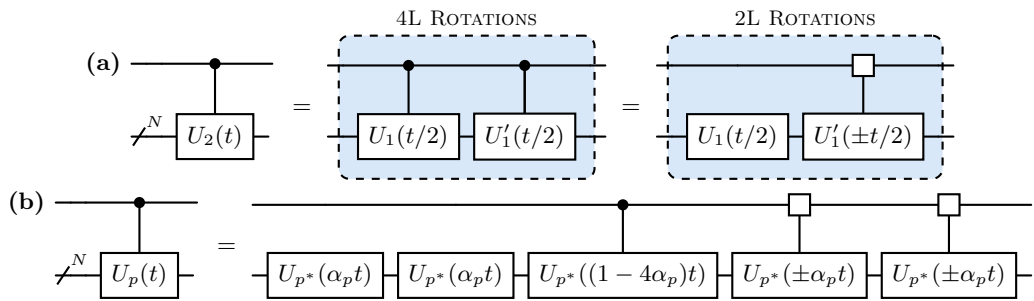

For certain applications - such as estimating properties of arbitrary quantum states - directional control may result in undesired effects.
Instead, we might wish to explicitly apply the \textit{controlled} time evolution unitary.
Using the symmetry in the second-order Trotter decomposition (Eq. \ref{eq:second-order}), we can implement the controlled unitary without doubling the number of arbitrary rotations by directionally controlling half of the time evolutions (subfigure \ref{fig:improved-compilation}a).

In this circuit, $U_1$ is left \textit{uncontrolled} and each time evolution within $U_1^\prime$ is \textit{directionally-controlled}.
When the control qubit is in the $\ket{0}$ state, the time evolutions in $U_1^\prime$ will sequentially undo the time evolutions of the terms in $U_1$, resulting in the identity.

This method only uses $2L$ rotations to achieve the controlled second-order Trotter decomposition, as compared to the $4L$ rotations required by the naive construction.
Higher-order Trotter decompositions constructed recursively from the second-order decomposition (Eq. \ref{eq:higher-order}) can be controlled by controlling the component second-order decompositions, therefore requiring only $2L5^{(p/2)-1}$ rotations.

Given the structure in higher-order formulae, we can further reduce the number of CNOT gates required using similar techniques.
Shown in subfigure \ref{fig:improved-compilation}b, a controlled application of $U_p$ can be achieved by leaving the first two applications of $U_{p^*}(\alpha_{p} t)$ uncontrolled, controlling $U_{p^*}((1-4\alpha_{p}) t)$, and then directionlly controlling the final two applications of $U_{p^*}(\alpha_{p} t)$.
When the control qubit is in the $\ket{0}$ state, the reversed time evolutions will undo the prior uncontrolled unitaries.

\section{\label{sec:conclusions}Conclusion}

In a fault-tolerant quantum architecture, each arbitrary rotation consumes multiple magic states, making these operations expensive to perform.
Here, we show how to reduce the number of arbitrary rotations required to apply a controlled Trotterization by a factor of two. This means that controlled time evolution requires the same number of arbitrary rotations as uncontrolled time evolution.

The techniques that we describe are motivated by the assumption that arbitrary rotations of logical qubits require the most computational resources in quantum simulation algorithms. 
New techniques to approximate controlled, logical rotations or to prepare magic states may substantially reduce the cost of these rotations and make the techniques we describe less influential.
Additionally, quantum simulation algorithms do not always require controlled time evolution and instead may be more efficient by using the uncontrolled or directionally controlled unitaries.

The assumption that coherent control of logical quantum operations substantially increases the computational resources influences the design of quantum algorithms. 
In this work, we show that controlled time evolution can be implemented using essentially the same resources as uncontrolled time evolution. 
This emphasizes the need to fully compile circuits into standard gate sets and analyze different compilation techniques in order to appropriately estimate the cost of performing different logical operations. 
\\

\begin{acknowledgments}
% \noindent \textbf{Acknowledgments}

W.A.S. is supported by the Department of Defense (DOD) through the National Defense Science \& Engineering Graduate (NDSEG) Fellowship Program.
P.J.L. is supported by US DOE Grant DE-SC0023707 under the Office of Nuclear Physics Quantum Horizons program for the ``{\bf Nu}clei and {\bf Ha}drons with {\bf Q}uantum computers (NuHaQ)" project.
\end{acknowledgments}

\bibliography{citations}

\end{document}